\documentclass[%
 reprint,
nofootinbib,
 amsmath,amssymb,
 aps,
]{revtex4-2}
\usepackage{amsmath}
\usepackage[colorlinks=true, linkcolor=blue, citecolor=blue]{hyperref}
\usepackage{graphicx}
\usepackage{dcolumn}
\usepackage{bm}

\begin{document}

\preprint{APS/123-QED}

\title{A Unified Framework for Virtual Wave Transform: From Generalized Formulation to Excitation-Specific Projection}

\author{Pengfei Zhu}
 \email{pengfei.zhu@bam.de}
\affiliation{%
 Bundesanstalt für Materialforschung and -prüfung (BAM), 12205 Berlin, Germany
}%

\author{Julien Lecompagnon}
\affiliation{%
	Bundesanstalt für Materialforschung and -prüfung (BAM), 12205 Berlin, Germany
}%

\author{Philipp Daniel Hirsch}
\affiliation{%
	Bundesanstalt für Materialforschung and -prüfung (BAM), 12205 Berlin, Germany
}%

\author{Mathias Ziegler}
\affiliation{%
	Bundesanstalt für Materialforschung and -prüfung (BAM), 12205 Berlin, Germany
}%

\date{\today}

\begin{abstract}
  We present a unified theoretical framework for the mapping between diffusive and wave-like dynamics, formulated as a spectral integral operator acting on temporal fields. By introducing an analytic continuation in the complex frequency plane, we establish an explicit correspondence between thermal diffusion and a virtual wave field governed by a hyperbolic equation. This mapping is shown to define a causal, compact Fredholm operator that acts as a nonstationary low-pass filter, thereby revealing the intrinsic information loss of diffusive processes and the fundamental ill-posedness of the inverse reconstruction.
  Within this operator framework, we demonstrate that commonly used excitation schemes—including pulse, lock-in, chirped, and coded excitations—emerge as distinct projections onto subspaces of a single underlying transformation, corresponding to different sampling strategies of its spectral structure. This unifies previously disparate virtual wave formulations and provides a systematic interpretation of excitation design in terms of operator sampling and information encoding. The framework further generalizes to matrix-valued systems and suggests a spectral-geometric interpretation of temporal evolution across diffusive and propagative regimes.
\end{abstract}

\maketitle

\section{Introduction}
\begin{figure*}[t]
	\centering
	\includegraphics[width=\textwidth]{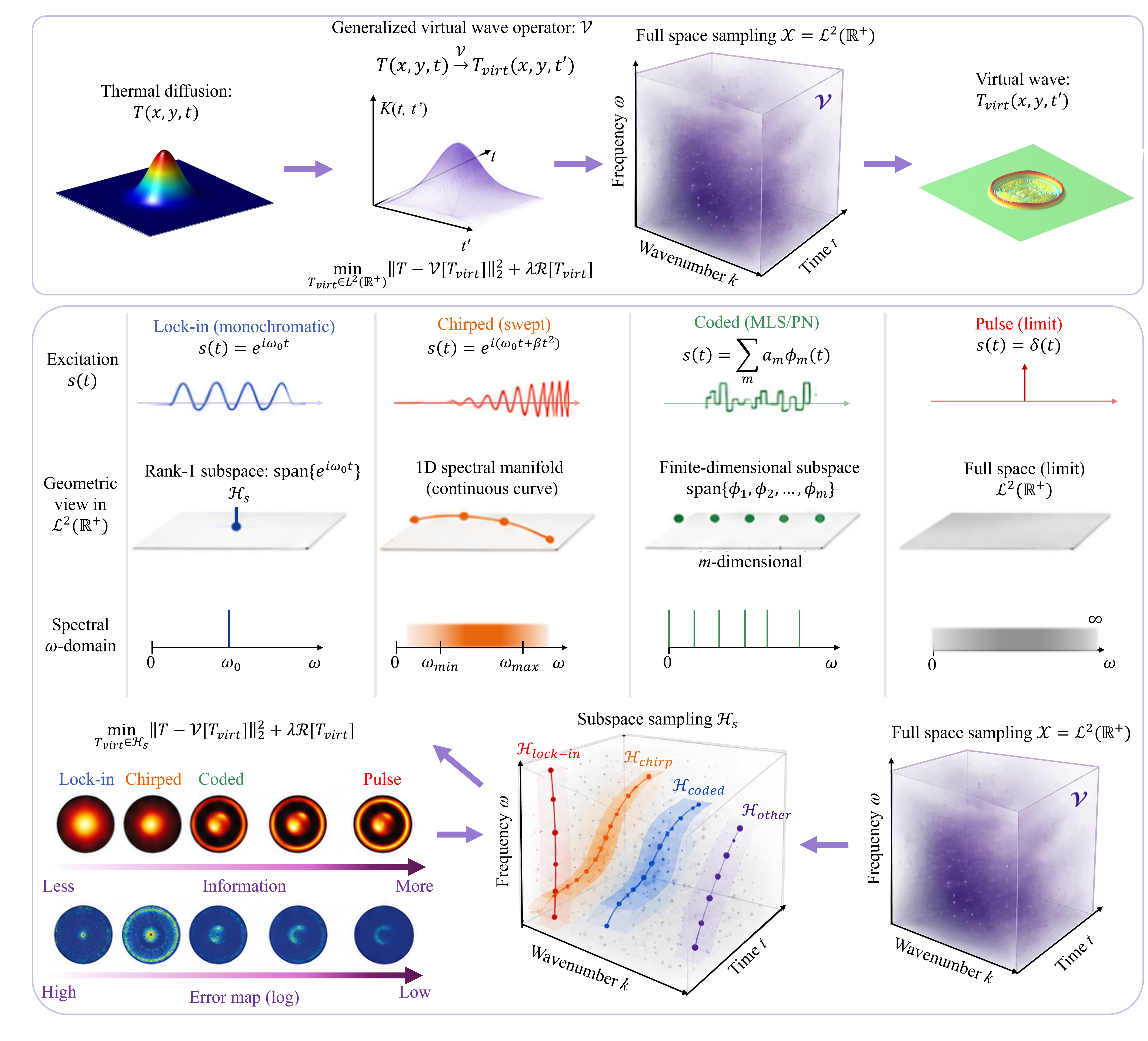}
	\caption{Spectral-geometric interpretation of generalized virtual-wave transformations.
		A generalized virtual-wave operator $\mathcal{V}$ maps thermal diffusion fields into wave-like representations through inverse reconstruction in a spectral function space. Upper panel: full-space sampling corresponds to reconstruction in the complete Hilbert space $\mathcal{X}=L^2(\mathbb{R}^+)$, yielding a virtual-wave field $T_{\mathrm{virt}}(x,y,t')$. Middle panel: different excitation schemes define distinct subspace samplings in spectral space, including monochromatic lock-in excitation (rank-1 subspace), chirped excitation (continuous one-dimensional spectral manifold), coded excitation (finite-dimensional subspace), and pulse excitation as the full-space limit. Lower panel: reconstruction quality and information recovery improve as the sampled spectral subspace expands toward full-space coverage. The geometric interpretation illustrates excitation-dependent trajectories in the joint $(k,\omega,t)$ spectral manifold and their relation to the complete operator space.}\label{fig1}
\end{figure*}

The classical heat diffusion equation is a macroscopic phenomenological model derived under the assumption of instantaneous heat flux response (Fourier's law)~\cite{1}. As a consequence, it predicts infinite propagation speed of thermal disturbances, which formally violates relativistic causality. Despite these fundamental limitations, the classical heat diffusion equation remains of profound practical importance in engineering~\cite{2} and applied physics~\cite{3}. Its macroscopic and phenomenological nature enables a simple yet effective description of heat transport in a wide range of materials~\cite{4} and conditions~\cite{5}. In particular, diffusion-based thermal models have been extensively employed for the characterization of thermophysical properties, such as thermal diffusivity~\cite{6,7} and conductivity~\cite{8,9}, as well as in various non-destructive evaluation (NDE) techniques~\cite{10,11}. In such approaches, modulated~\cite{12} or pulsed~\cite{13} heating gives rise to so-called diffusion-wave fields~\cite{14}, whose amplitude and phase encode valuable information about internal structures~\cite{15}. This duality, mathematical non-causality versus practical effectiveness, suggests that while the diffusion equation is sufficient for forward modeling~\cite{16}, additional considerations are required for accurate inversion~\cite{17}, reconstruction~\cite{18}, and interpretation of thermal signals~\cite{19}. In addition, due to its parabolic nature, thermal disturbances governed by diffusion dynamics do not exhibit a finite propagation front~\cite{20,21}. Instead, any localized excitation instantaneously spreads over the entire spatial domain with exponentially decaying amplitude, This absence of a well-defined wavefront fundamentally distinguishes diffusive transport from hyperbolic wave phenomena~\cite{22}. As a result, key concepts that are central in wave physics—such as travel time~\cite{23}, front tracking~\cite{24}, and ray-based interpretation~\cite{25}—are no longer applicable in a strict sense. This lack of causal propagation structure complicates the physical interpretation of measured thermal signals, especially in inverse problems where spatial localization and defect reconstruction are required~\cite{26,27}. In particular, the smearing effect inherent in diffusion processes leads to ill-posedness and reduced spatial resolution in thermal imaging and non-destructive evaluation~\cite{28,29,30,31}. Recently, the concept of virtual wave was proposed~\cite{32,33,34,35}, which is promising to solve the previously mentioned fundamental problem in thermal diffusion-wave fields. The concept of virtual waves originated from attempts to overcome information loss and resolution limits caused by irreversible processes such as acoustic attenuation and heat diffusion in imaging. Initially emerging in photoacoustic and thermographic reconstruction, it formalizes an idealized, time-reversible wave signal that satisfies the wave equation and preserves information, in contrast to real diffusive or dissipative signals. This idea was mathematically grounded through inverse-problem theory (notably Romanov’s framework), leading to the formulation of a local transformation—expressed as a Fredholm integral equation—that links measured diffusive fields to their propagative “virtual” counterparts. The initial contribution of applying the virtual wave or equivalent wave field transform to heat diffusion problem were developed by Burgholzer and Gershenson et al~\cite{36,37}. However, Gershenson’s derivation is essentially an ad hoc Laplace-based construction of a transform that mimics wave behavior, lacking the rigorous inverse-problem foundation and uniqueness guarantees provided in Burgholzer’s Fredholm-integral formulation. The virtual wave method proposed by  Burgholzer et al. is only suitable for Dirac pulse-based heat diffusion problem. Then, Zhu et al. define this kind of transform as "special virtual wave transform" and propose the generalized formalism, i.e., generalized virtual wave transform~\cite{38,39}. Here, we construct a mathematically defined mapping from diffusive heat fields to virtual wave fields via an analytic continuation in frequency space, and reformulate it as a unified Fredholm operator framework in which different excitation schemes are interpreted as distinct sampling strategies of the same underlying transformation.

The heat conduction in isotropic medium is governed by the heat diffusion equation:
\begin{equation}
	\left(\nabla^2 - \frac{1}{\alpha}\frac{\partial}{\partial t}\right)
	T(\mathbf{r}, t)
	= -\frac{1}{\kappa} Q(\mathbf{r}, t),
	\label{eq1}
\end{equation}
where $T(\mathbf{r}, t)$ denotes the temperature field, $\alpha$ denotes the thermal diffusivity, $\kappa$ denotes the thermal conductivity, and $Q(\mathbf{r}, t)$ denotes an externally applied heat source. Equation~\eqref{eq1} is obtained by rearranging the standard heat equation into a Helmholtz-like operator form for later comparison.

A scalar wave field $p(\mathbf{r}, t)$ obeys the wave equation:
\begin{equation}
	\left(\nabla^2 - \frac{1}{c^2}\frac{\partial^2}{\partial t^2}\right)
	p(\mathbf{r}, t)
	= -\frac{1}{c^2} Q(\mathbf{r}, t),
	\label{eq2}
\end{equation}
where $c$ is the wave speed. We assume the same temporal waveform of the source term $Q(\mathbf{r}, t)$ in both equations, while allowing different physical units.

To establish a formal connection between diffusion and wave phenomena, we introduce a virtual wave field $T_{\mathrm{virt}}(\mathbf{r}, t)$ defined as:
\begin{equation}
	\left(\nabla^2 - \frac{1}{c^2}\frac{\partial^2}{\partial t^2}\right)
	T_{\mathrm{virt}}(\mathbf{r}, t)
	= -\frac{1}{c^2} Q(\mathbf{r}, t).
	\label{eq3}
\end{equation}

The Fourier transform pair is defined as:
\begin{equation}
	\tilde{T}(\mathbf{r}, \omega)
	= \int_{-\infty}^{\infty} T(\mathbf{r}, t)e^{-i\omega t}\, dt,
	\label{eq4}
\end{equation}

\begin{equation}
	T(\mathbf{r}, t)
	= \frac{1}{2\pi} \int_{-\infty}^{\infty}
	\tilde{T}(\mathbf{r}, \omega)e^{i\omega t}\, d\omega,
	\label{eq5}
\end{equation}
where $\tilde{T}(\mathbf{r}, \omega)$ is the temperature field in frequency space.

Applying the Fourier transform to Eq.~\eqref{eq1} yields:
\begin{equation}
	\left(\nabla^2 - \sigma^2(\omega)\right)\tilde{T}(\mathbf{r}, \omega)
	= -\frac{1}{\kappa}\tilde{Q}(\mathbf{r}, \omega),
	\label{eq6}
\end{equation}
where $\sigma^2(\omega) \equiv i\omega/\alpha$. The complex wavenumber $\sigma(\omega)$ is chosen such that $\mathrm{Re}[\sigma(\omega)] > 0$, ensuring spatially decaying thermal diffusion waves.

Similarly, the frequency-domain form of Eq.~\eqref{eq3} reads:
\begin{equation}
	\left(\nabla^2 + k^2(\omega)\right)\tilde{T}_{\mathrm{virt}}(\mathbf{r}, \omega)
	= -\frac{1}{c^2}\tilde{Q}(\mathbf{r}, \omega),
	\label{eq7}
\end{equation}
with the real wavenumber $k(\omega) \equiv \omega/c$.
Eqs.~\eqref{eq6} and \eqref{eq7} are both Helmholtz-type equations; however, Eq.~\eqref{eq6} involves a complex wavenumber characteristic of diffusion, while Eq.~\eqref{eq7} describes propagating waves.

Motivated by the formal similarity between Eqs.~\eqref{eq6} and \eqref{eq7}, we introduce a correspondence between the thermal diffusion field and the virtual wave field via analytic continuation in the complex frequency plane. Specifically, replacing $\omega$ in Eq.~\eqref{eq7} by $-ic\sigma(\omega)$ leads to
\begin{equation}
	\tilde{T}(\mathbf{r}, \omega)
	= \frac{c^2}{\kappa}\,
	\tilde{T}_{\mathrm{virt}}(\mathbf{r}, -ic\sigma(\omega)),
	\label{eq8}
\end{equation}
which defines a mapping between the diffusion field and the virtual wave field in the frequency domain. The rigorous justification of this analytic continuation is provided in the Appendix~\ref{app:subsecA1}

Equation~\eqref{eq8} should be regarded as a mathematically constructed correspondence rather than a physical equivalence. Transforming Eq.~\eqref{eq8} back to the time domain yields
\begin{equation}
	T(\mathbf{r}, t)
	= \frac{1}{2\pi} \int_{-\infty}^{\infty}
	\frac{c^2}{\kappa}
	\tilde{T}_{\mathrm{virt}}(\mathbf{r}, -ic\sigma(\omega))
	e^{i\omega t} \, d\omega,
	\label{eq9}
\end{equation}
where
\begin{equation}
	\tilde{T}_{\mathrm{virt}}(\mathbf{r}, -ic\sigma(\omega))
	= \int_{-\infty}^{\infty}
	T_{\mathrm{virt}}(\mathbf{r}, t')
	e^{-c\sigma(\omega)t'} \, dt'.
	\label{eq10}
\end{equation}

Equation~\eqref{eq9} can be rewritten as a Fredholm integral equation of the first kind,
\begin{equation}
	T(\mathbf{r}, t)
	= \int_{-\infty}^{\infty}
	T_{\mathrm{virt}}(\mathbf{r}, t')\, K(t, t') \, dt',
	\label{eq11}
\end{equation}
where the kernel $K(t,t')$ enforces causality. The kernel can be evaluated analytically as
\begin{equation}
	K(t, t')
	= \frac{c^2}{\kappa}
	\frac{t'}{2\sqrt{\pi}(\alpha t)^{3/2}}
	\exp\!\left(-\frac{c^2 t'^2}{4\alpha t}\right)
	H(t)\,H(t'),
	\label{eq12}
\end{equation}
where $H(\cdot)$ denotes the Heaviside step function.

The kernel explicitly enforces causality and reflects the irreversible nature of thermal diffusion, whereby the temperature field at time $t$ depends on contributions from all prior virtual-wave times $t' > 0$.
Eqs.~\eqref{eq11} and ~\eqref{eq12} connect the temperature signal to the virtual wave signal at the same spatial location $\mathbf{r}$ via a Fredholm integral equation of the first kind (Fig.~\ref{fig1}).

\section{Excitation-Induced Subspace Sampling of the Generalized Virtual Wave Transform}
\begin{figure*}[t]
	\centering
	\includegraphics[width=\textwidth]{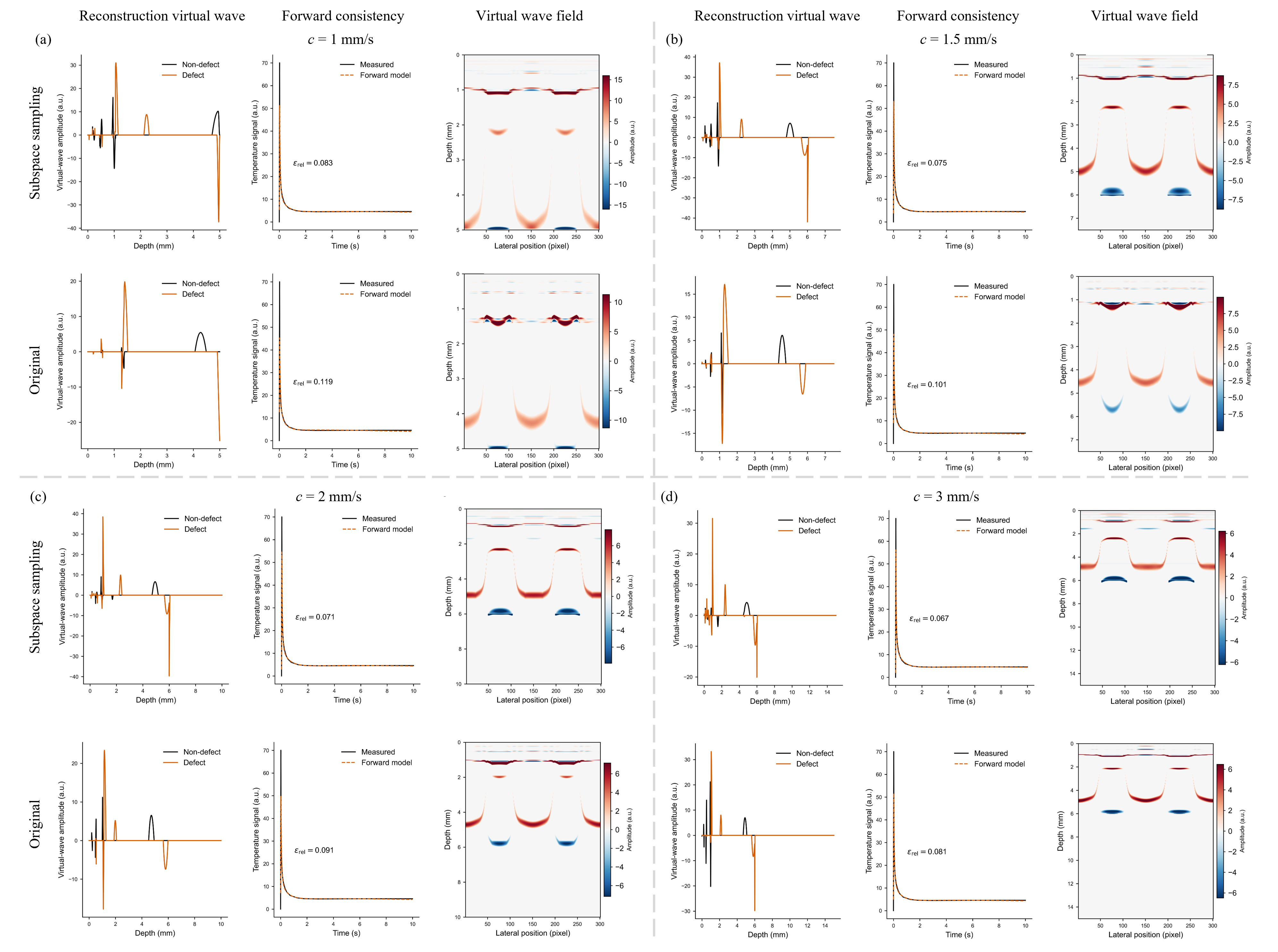}
	\caption{Dependence of virtual wave reconstruction on the assumed propagation velocity $c$. Columns show reconstructed waveforms, forward consistency, and final defect images; top/bottom rows compare subspace sampling and unconstrained inversion. Subspace sampling yields stable structures and low error across $c=1–3$ mm/s, whereas the unconstrained case exhibits strong sensitivity, with distorted waveforms and degraded imaging.}\label{fig2}
\end{figure*}
Thermal diffusion intrinsically suppresses high-frequency information through exponential spectral attenuation, leading to severe ill-posedness in inverse reconstruction problems. Conventional virtual wave approaches partially alleviate this limitation by introducing wave-like representations of diffusive transport. However, existing formulations are typically excitation-specific and lack a unified operator interpretation. In this work, we show that a broad class of virtual-wave-based excitation schemes can be understood within a single generalized operator framework, where different excitations correspond to distinct subspace projections or sampling strategies of the same generalized virtual wave transform (GVWT).

The diffusion field is governed by
\begin{equation}
	\frac{\partial T(\mathbf r,t)}{\partial t}
	=
	\alpha \nabla^2 T(\mathbf r,t),
\end{equation}
where $\alpha$ denotes the thermal diffusivity. Introducing a virtual wave field $T_{\mathrm{virt}}(\mathbf r,t)$, the diffusion process can be represented through a generalized virtual wave operator
\begin{equation}
	T(\mathbf r,t)
	=
	\mathcal V[T_{\mathrm{virt}}](\mathbf r,t)
	=
	\int_0^\infty
	K(t,t')
	T_{\mathrm{virt}}(\mathbf r,t')
	dt',
	\label{eq_main_gvwt}
\end{equation}
where $K(t,t')$ denotes the diffusion-wave kernel. The operator $\mathcal V$ acts as a causal smoothing transformation that maps propagative virtual-wave dynamics into exponentially damped diffusion fields.

In practice, the virtual wave field is not arbitrary, but is generated by a prescribed excitation waveform $s(t)$. Let
\begin{equation}
	Q(\mathbf r,t)
	=
	q(\mathbf r)s(t),
\end{equation}
then the virtual wave field can be expressed as
\begin{equation}
	T_{\mathrm{virt}}(\mathbf r,t)
	=
	\mathcal W[s]
	=
	\int_0^\infty
	G_{\mathrm{wave}}(\mathbf r,t-\tau)
	s(\tau)d\tau,
\end{equation}
where $\mathcal W$ denotes the wave propagation operator associated with the wave Green's function $G_{\mathrm{wave}}$.

Consequently, the measured temperature field is governed by the composite operator
\begin{equation}
	\mathcal A
	=
	\mathcal V
	\circ
	\mathcal W,
\end{equation}
such that
\begin{equation}
	T(\mathbf r,t)
	=
	(\mathcal V\circ\mathcal W)[s].
	\label{eq_main_operator}
\end{equation}
\begin{figure}[t]
	\centering
	\includegraphics[width=\columnwidth]{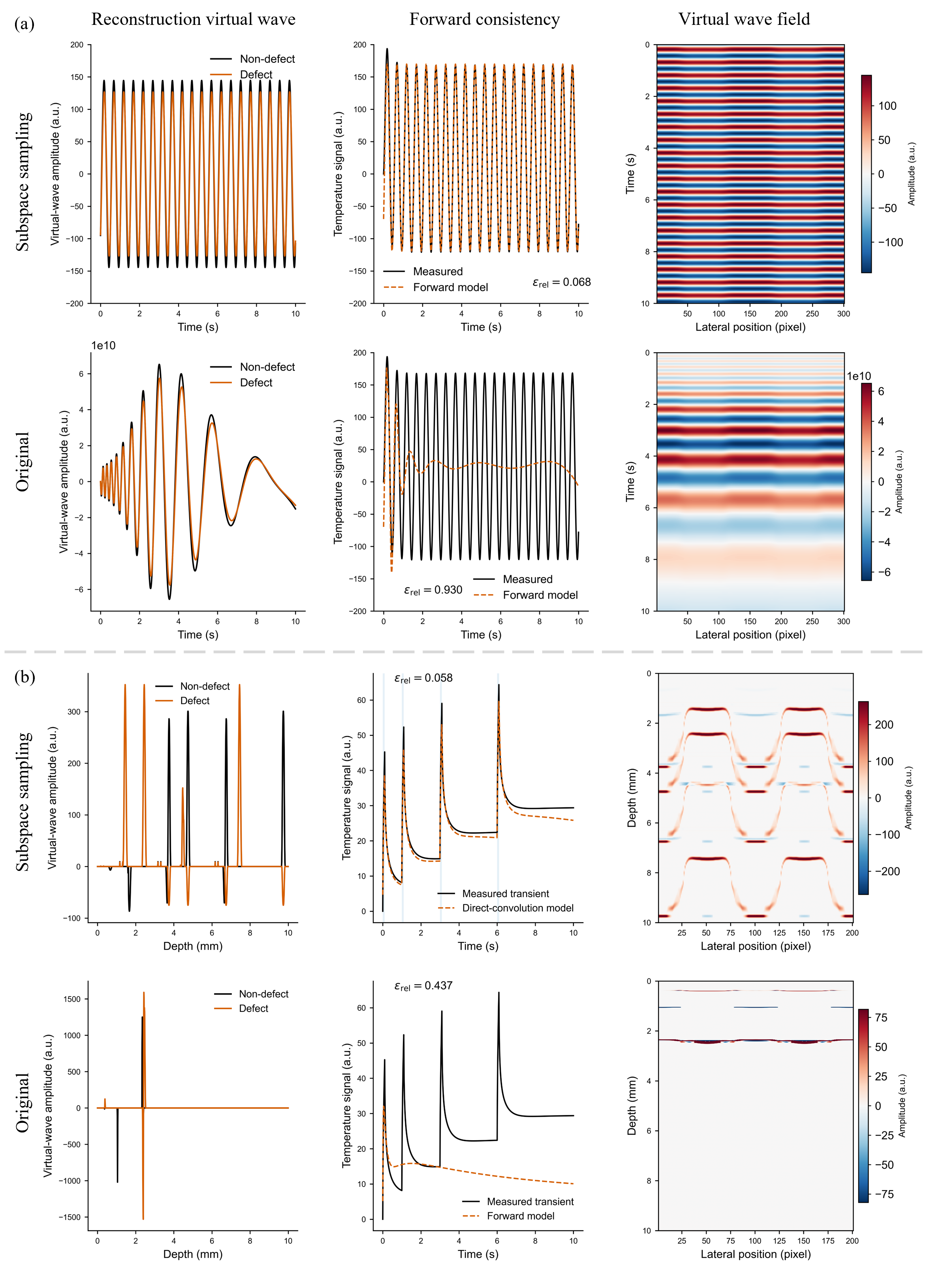}
	\caption{Virtual wave reconstruction under lock-in (a) and chirp-pulsed (b) excitation. Columns show reconstructed waveforms, forward consistency, and resulting wave fields (or defect images). Subspace sampling (top) yields stable, physically consistent reconstructions with low error, whereas the unconstrained inversion (bottom) exhibits strong distortion, poor consistency, and degraded imaging performance.}\label{fig3}
\end{figure}
This formulation reveals that excitation design is fundamentally equivalent to selecting admissible subspaces of the generalized virtual-wave operator. Since the excitation waveform constrains the temporal modes that can be generated through $\mathcal W$, the virtual wave field no longer spans the full infinite-dimensional space $L^2(\mathbb R^+)$, but instead belongs to an excitation-induced subspace
\begin{equation}
	\mathcal H_s
	\subset
	L^2(\mathbb R^+).
\end{equation}

Different excitation schemes therefore correspond to different subspace projections or sampling strategies of the same generalized operator framework.

For monochromatic lock-in excitation,
\begin{equation}
	s(t)=e^{i\omega_0 t},
\end{equation}
the admissible solution space collapses into the rank-1 harmonic subspace
\begin{equation}
	\mathcal H_{\omega_0}
	=
	\mathrm{span}
	\left\{
	e^{i\omega_0 t}
	\right\}.
\end{equation}
The inversion problem is therefore strongly regularized, since only a single spectral mode is sampled.

For chirped excitation,
\begin{equation}
	s(t)
	=
	e^{i(\omega_0 t+\beta t^2)},
\end{equation}
the instantaneous frequency
\begin{equation}
	\omega_{\mathrm{inst}}(t)
	=
	\omega_0+2\beta t
\end{equation}
defines a continuous spectral trajectory in frequency space. In this case, the excitation probes the operator along a continuous low-dimensional manifold, increasing spectral coverage while maintaining partial regularization.

For coded excitation,
\begin{equation}
	s(t)
	=
	\sum_{m}
	a_m\phi_m(t),
\end{equation}
the admissible virtual-wave field belongs to the finite-dimensional subspace
\begin{equation}
	\mathcal H_{\mathrm{coded}}
	=
	\mathrm{span}
	\left\{
	\phi_m(t)
	\right\}_{m=1}^{M},
\end{equation}
where the dimensionality is determined by the number of coding modes $M$. Such excitation schemes enable multiplexed sampling of the operator spectrum and increase information capacity.

For pseudo-noise (PN) excitation,
\begin{equation}
	s(t)
	=
	\sum_n
	s[n]\delta(t-n\Delta t),
\end{equation}
the excitation induces a discrete sampling of shifted kernel components,
\begin{equation}
	T(t)
	=
	\sum_n
	s[n]K(t,n\Delta t),
\end{equation}
which corresponds to high-dimensional discrete sampling of the generalized operator manifold.

These excitation-dependent subspaces reveal a fundamental trade-off between inversion stability and information capacity. Low-dimensional excitation subspaces provide strong regularization and robust inversion, but restrict spectral information content. In contrast, high-dimensional sampling strategies increase recoverable information while simultaneously amplifying ill-posedness and noise sensitivity. This relationship can be qualitatively expressed as
\begin{equation}
	\mathcal C_{\mathrm{inv}}
	\sim
	f\!\left(
	\dim(\mathcal H_s)
	\right),
\end{equation}
where $\mathcal C_{\mathrm{inv}}$ denotes the effective inversion complexity and $\dim(\mathcal H_s)$ denotes the dimensionality of the excitation-induced subspace.
\begin{table}[t]
	\centering
	\caption{Pulse-Subspace GVWT via ADMM Reconstruction}
	\label{tab:subspace_algorithm}
	\renewcommand{\arraystretch}{1.25}
	\small
	\begin{tabular}{@{}l@{}}
		\hline
		
		\textbf{Input:}
		$\mathbf T,\mathbf K,\lambda,\rho,M,\sigma$ \\
		
		\hline
		
		\textbf{Initialization:}
		$\mathbf a^{(0)}=\mathbf0$,
		$\mathbf z^{(0)}=\mathbf0$,
		$\mathbf u^{(0)}=\mathbf0$,
		$k=0$ \\
		
		\textbf{Pulse basis:}
		$\mathbf\Phi=[\phi_1,\phi_2,\ldots,\phi_M]$ \\
		
		\textbf{Reduced operator:}
		$\mathbf K_s=\mathbf K\mathbf\Phi$ \\
		
		\textbf{Forward model:}
		$\mathbf T=\mathbf K_s\mathbf a$ \\
		
		\textbf{while} not converged \textbf{do} \\
		
		\quad\begin{tabular}{@{}l@{}}
			
			\quad\hspace{3pt}\vrule\hspace{6pt}
			\textbf{(A) Coefficient update} \\
			
			\quad\hspace{3pt}\vrule\hspace{6pt}
			$\mathbf a^{(k+1)}
			=
			(\mathbf K_s^{T}\mathbf K_s+\rho\mathbf I)^{-1}$ \\
			
			\quad\hspace{3pt}\vrule\hspace{6pt}
			$\times
			\left[
			\mathbf K_s^{T}\mathbf T
			+
			\rho(\mathbf z^{(k)}-\mathbf u^{(k)})
			\right]$ \\
			
			\quad\hspace{3pt}\vrule\hspace{6pt}
			\textbf{(B) Soft-thresholding} \\
			
			\quad\hspace{3pt}\vrule\hspace{6pt}
			$\mathbf z^{(k+1)}
			=
			\mathcal S_{\lambda/\rho}
			\left(
			\mathbf a^{(k+1)}
			+
			\mathbf u^{(k)}
			\right)$ \\
			
			\quad\hspace{3pt}\vrule\hspace{6pt}
			\textbf{(C) Dual update} \\
			
			\quad\hspace{3pt}\vrule\hspace{6pt}
			$\mathbf u^{(k+1)}
			=
			\mathbf u^{(k)}
			+
			\mathbf a^{(k+1)}
			-
			\mathbf z^{(k+1)}$ \\
			
			\quad\hspace{3pt}\vrule\hspace{6pt}
			$k\leftarrow k+1$ \\
			
		\end{tabular} \\
		
		\textbf{end while} \\
		
		\textbf{Virtual-wave reconstruction:}
		$\mathbf T_{\mathrm{virt}}=\mathbf\Phi\mathbf a$ \\
		
		\textbf{Output:}
		$\mathbf T_{\mathrm{virt}}$ \\
		
		\hline
	\end{tabular}
\end{table}
Within this perspective, conventional pulsed, lock-in, chirped, coded, and PN-based thermography schemes no longer represent fundamentally distinct virtual-wave transforms. Instead, they emerge naturally as different degeneracy limits, projections, or sampling strategies of the same generalized virtual wave operator. Excitation design therefore becomes equivalent to controlling the geometry of the accessible operator subspace, providing a unified framework for analyzing spectral observability, inversion stability, and information capacity in diffusion-wave imaging systems.

Figure~\ref{fig1} summarizes the central concept of the proposed framework from a spectral-geometric perspective. Instead of treating lock-in, chirped, coded, and pulsed thermography as fundamentally different modalities, the framework interprets them as different sampling strategies of the same generalized virtual-wave operator in the Hilbert space $L^2(\mathbb{R}^+)$. Each excitation waveform defines a specific spectral subspace $\mathcal{H}_s$, ranging from low-dimensional manifolds for monochromatic excitation to near-complete spectral coverage for pulsed excitation. Under this interpretation, the reconstruction problem becomes a subspace-limited inverse mapping between diffusive and wave-like dynamics. The figure further illustrates that increasing spectral coverage systematically enhances recoverable information and reduces reconstruction error, ultimately approaching the ideal full-space virtual-wave reconstruction limit.
\begin{table}[t]
	\centering
	\caption{Lock-in Subspace GVWT via L2-Smooth Reconstruction}
	\label{tab:lockin_algorithm}
	\renewcommand{\arraystretch}{1.25}
	\small
	\begin{tabular}{@{}l@{}}
		\hline
		
		\textbf{Input:}
		$\mathbf T,\mathbf K,f_0,\lambda_1,\lambda_2$
		\\
		
		\hline
		
		\textbf{Lock-in projection:}
		$\mathbf T \rightarrow \mathbf T_{\rm lockin}$
		\\
		
		\textbf{Lock-in basis:}
		$\mathbf\Phi=[\phi_1,\phi_2,\ldots,\phi_M]$
		\\
		
		$\phi_m(\tau)
		=
		g_m(\tau)\cos(2\pi f_0\tau)$
		\\
		
		$\phi_{m+M}(\tau)
		=
		g_m(\tau)\sin(2\pi f_0\tau)$
		\\
		
		\textbf{Reduced operator:}
		$\mathbf K_s=\mathbf K\mathbf\Phi$
		\\
		
		\textbf{Forward model:}
		$\mathbf T_{\rm lockin}
		=
		\mathbf K_s\mathbf a$
		\\
		
		\textbf{Coefficient reconstruction:}
		\\
		
		\quad\begin{tabular}{@{}l@{}}
			
			\quad\hspace{3pt}\vrule\hspace{6pt}
			Solve
			\\
			
			\quad\hspace{3pt}\vrule\hspace{6pt}
			$\displaystyle
			\mathbf a
			=
			\arg\min_{\mathbf a}
			\|
			\mathbf K_s\mathbf a
			-
			\mathbf T_{\rm lockin}
			\|_2^2$
			\\
			
			\quad\hspace{3pt}\vrule\hspace{6pt}
			$\displaystyle
			+
			\lambda_1
			\|\mathbf a\|_2^2
			+
			\lambda_2
			\|\mathbf D\mathbf a\|_2^2$
			\\
			
			\quad\hspace{3pt}\vrule\hspace{6pt}
			$\displaystyle
			\mathbf a
			=
			(
			\mathbf K_s^T\mathbf K_s
			+
			\lambda_1\mathbf I
			+
			\lambda_2\mathbf D^T\mathbf D
			)^{-1}$
			\\
			
			\quad\hspace{3pt}\vrule\hspace{6pt}
			$\displaystyle
			\mathbf K_s^T
			\mathbf T_{\rm lockin}$
			\\
			
		\end{tabular}
		\\
		
		\textbf{Virtual-wave reconstruction:}
		$\mathbf T_{\rm virt}
		=
		\mathbf\Phi\mathbf a$
		\\
		
		\textbf{Output:}
		$\mathbf T_{\rm virt}$
		\\
		
		\hline
	\end{tabular}
\end{table}
\section{Simulation Validation}
\begin{figure}[t]
	\centering
	\includegraphics[width=\columnwidth]{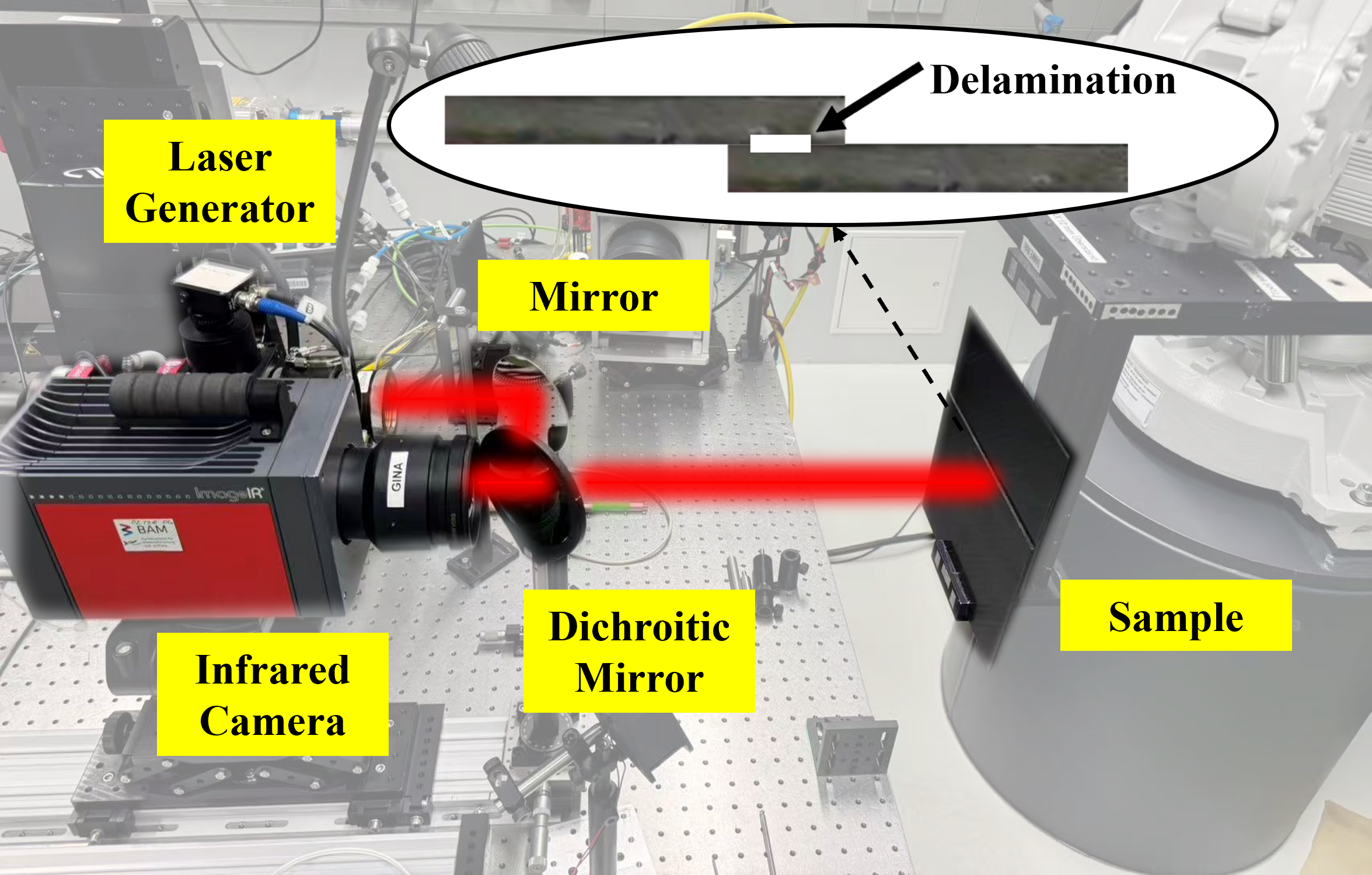}
	\caption{Photograph of photothermal experiments.}\label{fig4}
\end{figure}
Figure~\ref{fig2} reveals a pronounced contrast between the subspace-sampled reconstruction and the original virtual wave inversion when varying the assumed virtual wave velocity $c$. In the subspace sampling framework, the reconstructed virtual-wave signals and the corresponding defect images remain qualitatively stable over a wide range of $c$: the echo positions are preserved, and the defect morphology exhibits only minor amplitude variations. This behavior is consistent with the forward-consistency results, indicating that the inversion remains well-conditioned under velocity perturbations.
In contrast, the original virtual wave reconstruction (without subspace constraint) exhibits a strong and systematic dependence on $c$. As shown in the lower panels of Fig.~\ref{fig2}, changing the virtual wave speed leads to a clear drift of the reconstructed echoes in depth, accompanied by significant waveform distortion. More critically, for larger values of $c$, some physically expected reflections are progressively attenuated or even completely suppressed, resulting in partial or missing defect signatures in the final images. This effect directly demonstrates that the conventional formulation lacks robustness with respect to the spectral reparameterization inherent in the virtual wave mapping.
The underlying reason is that, in the absence of subspace restriction, the inversion operates in a fully unconstrained infinite-dimensional space, where the analytic continuation controlled by $c$ effectively redistributes spectral energy without any structural constraint. Consequently, small mismatches in $c$ translate into large variations in the reconstructed temporal modes, manifesting as echo drift and disappearance. By contrast, the excitation-induced subspace sampling acts as an intrinsic regularization, confining the solution to a physically admissible manifold and thereby stabilizing the reconstruction against such spectral distortions.
Table~\ref{tab:subspace_algorithm} summarizes the corresponding ADMM-based implementation of the subspace-constrained inversion, which enforces this low-dimensional structure in practice.
Overall, Fig.~\ref{fig2} highlights that the apparent “velocity sensitivity” of virtual wave reconstruction is not a benign parameter dependence but a direct signature of ill-posedness in the unconstrained inverse problem, which is effectively mitigated by subspace sampling.
\begin{table}[!t]
	\centering
	\caption{Coded-Excitation Subspace GVWT via ADMM Reconstruction}
	\label{tab:coded_algorithm}
	\renewcommand{\arraystretch}{1.35}
	\footnotesize
	
	\begin{tabular}{l}
		\hline
		
		\textbf{Input:}
		$\mathbf T,
		\mathbf K,
		\mathbf s,
		\lambda,
		\rho$
		\\
		
		\hline
		
		\textbf{Initialization:}
		
		$\mathbf a^{(0)}=\mathbf0$,
		$\mathbf z^{(0)}=\mathbf0$,
		$\mathbf u^{(0)}=\mathbf0$
		\\
		
		\textbf{Coding Sequence:}
		
		Construct excitation sequence
		$\mathbf s$
		\\
		
		\textbf{Coding Operator:}
		
		$\mathbf C=
		{\rm Toeplitz}(\mathbf s)$
		\\
		
		\textbf{Encoded Forward Kernel:}
		
		$\mathbf K_{\rm code}
		=
		\mathbf C\mathbf K$
		\\
		
		\textbf{Subspace Construction:}
		
		$\mathbf\Phi=
		[\phi_1,\phi_2,\ldots,\phi_M]$
		\\
		
		\textbf{Reduced Operator:}
		
		$\mathbf K_s
		=
		\mathbf K_{\rm code}
		\mathbf\Phi$
		\\
		
		\textbf{Forward Model:}
		
		$\mathbf T
		=
		\mathbf K_s\mathbf a$
		\\
		
		\textbf{while} not converged \textbf{do}
		\\
		
		\quad
		\begin{tabular}{@{}l@{}}
			
			\quad\hspace{3pt}\vrule\hspace{6pt}
			\textbf{(A) Coefficient update}
			\\
			
			\quad\hspace{3pt}\vrule\hspace{6pt}
			Solve
			\\
			
			\quad\hspace{3pt}\vrule\hspace{6pt}
			$\mathbf a^{(k+1)}
			=
			\arg\min_{\mathbf a}
			\|
			\mathbf K_s\mathbf a
			-
			\mathbf T
			\|_2^2
			+
			\lambda
			\|\mathbf a\|_1$
			\\
			
			\quad\hspace{3pt}\vrule\hspace{6pt}
			\textbf{(B) Soft-thresholding}
			\\
			
			\quad\hspace{3pt}\vrule\hspace{6pt}
			$\mathbf z^{(k+1)}
			=
			\mathcal S_{\lambda/\rho}
			(
			\mathbf a^{(k+1)}
			+
			\mathbf u^{(k)}
			)$
			\\
			
			\quad\hspace{3pt}\vrule\hspace{6pt}
			\textbf{(C) Dual update}
			\\
			
			\quad\hspace{3pt}\vrule\hspace{6pt}
			$\mathbf u^{(k+1)}
			=
			\mathbf u^{(k)}
			+
			\mathbf a^{(k+1)}
			-
			\mathbf z^{(k+1)}$
			\\
			
		\end{tabular}
		\\
		
		\textbf{end while}
		\\
		
		\textbf{Virtual-Wave Reconstruction:}
		
		$\mathbf T_{\rm virt}
		=
		\mathbf\Phi\mathbf a$
		\\
		
		\textbf{Pulse Compression:}
		
		$\mathbf T_{\rm comp}
		=
		\mathbf T_{\rm virt}
		\star
		\mathbf s(-t)$
		\\
		
		\textbf{Output:}
		
		$\mathbf T_{\rm comp}$
		\\
		
		\hline
	\end{tabular}
\end{table}

Tables~\ref{tab:lockin_algorithm} and~\ref{tab:coded_algorithm} summarize the forward-consistency errors and reconstruction quality metrics for lock-in and coded (chirp-pulsed) excitations, respectively, highlighting the consistently improved accuracy of the subspace-sampled approach over the unconstrained inversion.
Figure~\ref{fig3} shows that the advantage of subspace sampling persists across both harmonic (lock-in) and broadband (chirp-pulsed) excitation regimes. In the lock-in case, the reconstructed virtual wave remains strictly periodic and matches the measured signal with high fidelity, leading to a coherent wavefield, whereas the unconstrained inversion introduces amplitude distortion and loses temporal coherence. In the chirp-pulsed case, where the signal is inherently transient and spectrally rich, these differences become more pronounced: the subspace approach accurately captures discrete reflections and preserves defect localization, while the unconstrained method produces spurious structures and degraded images. This demonstrates that the subspace constraint effectively regularizes the inverse problem independently of the excitation type, ensuring stable waveform reconstruction and consistent defect imaging even under broadband excitation.
\begin{figure*}[t]
	\centering
	\includegraphics[width=\textwidth]{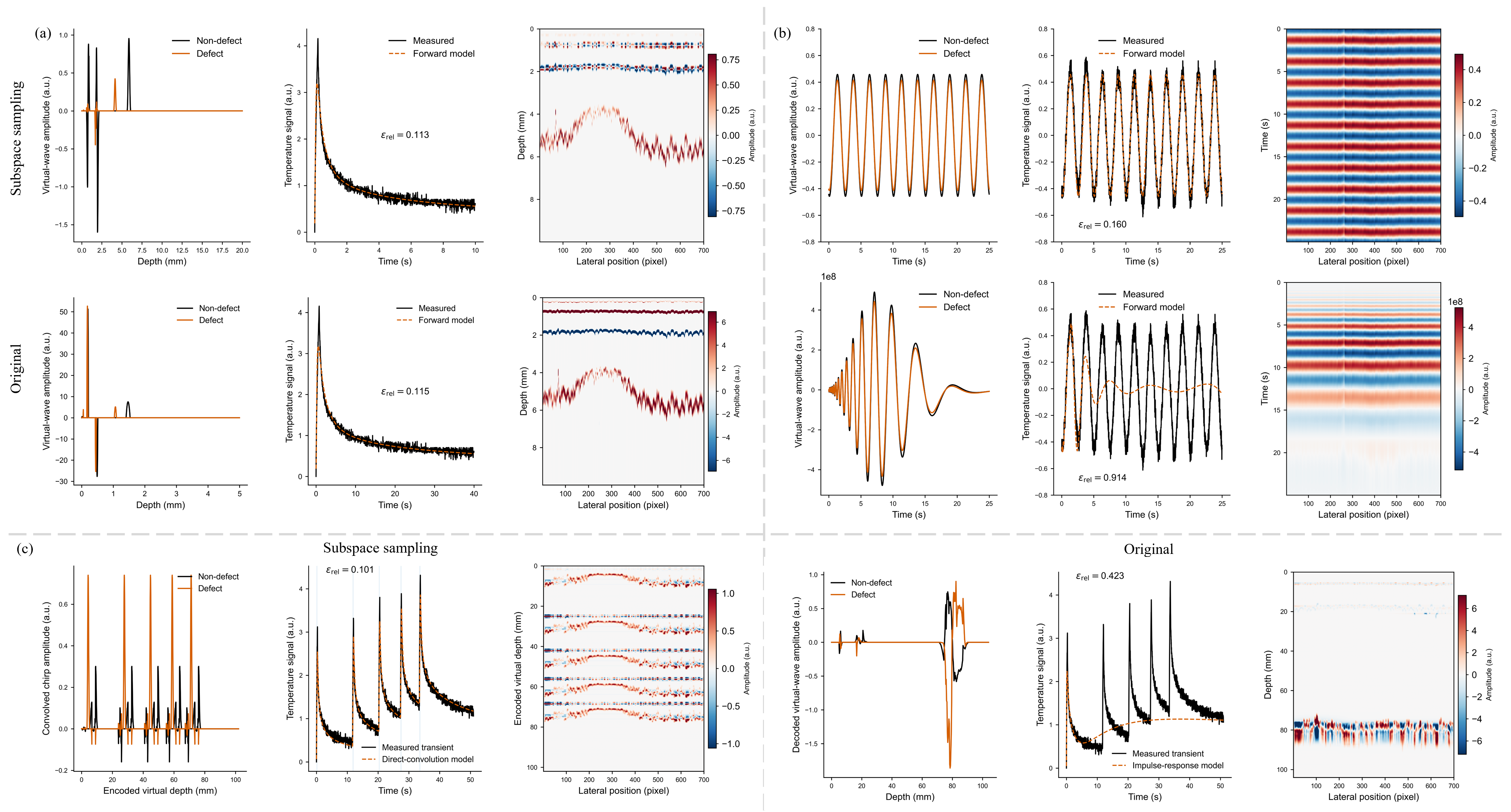}
	\caption{Virtual wave reconstruction for photothermal experimental  results: (a) Subspace sampling (Top) and original (bottom) reconstruction under pulse excitation. (b) Subspace sampling (Top) and original (bottom) reconstruction under lock-in excitation. (c) Subspace sampling (left) and original (right) reconstruction under chirp-pulsed excitation. Columns show reconstructed waveforms, forward consistency, and resulting wave fields (or defect images).}\label{fig5}
\end{figure*}
\section{Experimental Validation}
The experimental setup is shown in Fig.~\ref{fig4}. A laser generator serves as the excitation source, producing a collimated beam that is steered by a planar mirror and delivered to the specimen through a dichroic mirror. The dichroic element separates the excitation and detection paths, allowing the reflected signal to be recorded by an off-axis infrared camera (MWIR) equipped with imaging optics. All components are mounted on an optical breadboard and carefully aligned to ensure stable illumination and reproducible measurement conditions. The specimen is rigidly fixed on a holder to minimize mechanical vibrations during acquisition.
The sample consists of two bonded composite laminates containing an artificially introduced delamination, as indicated in Fig.~\ref{fig4}. The plate has a lateral dimension of approximately 225 mm, with each laminate having a thickness of 2 mm. The delamination is embedded beneath the surface and localized within the structure, forming a subsurface discontinuity. Such defects are representative of typical damage modes in composite materials, and they locally perturb the thermal/optical response, enabling detection with the present laser-based method.

Figure~\ref{fig5} highlights the excitation-dependent robustness of the generalized virtual-wave inversion and reveals a clear hierarchy of stability between subspace-constrained and unconstrained (“original”) reconstructions.
For pulsed excitation (Fig.~\ref{fig5}(a)), both subspace-sampled and original reconstructions yield nearly identical virtual-wave responses, forward consistency, and defect images. This agreement reflects the fact that pulsed excitation effectively approximates full-space spectral sampling, so that the inversion is weakly constrained by the excitation itself and does not strongly benefit from additional subspace regularization. Consequently, both approaches operate close to the ideal full-space limit, leading to comparable reconstruction quality. In contrast, for lock-in excitation (Fig.~\ref{fig5}(b)), the difference becomes pronounced. The subspace-sampled reconstruction accurately preserves the harmonic structure of the virtual wave and achieves good agreement with the forward model, whereas the original reconstruction exhibits clear amplitude distortion and degraded consistency (as indicated by the increased relative error). This behavior is a direct consequence of the rank-1 spectral projection imposed by lock-in excitation: without subspace constraint, the inversion is performed in an effectively over-parameterized space, leading to instability and loss of phase-coherent structure. The subspace formulation, by explicitly restricting the solution to the physically admissible harmonic manifold, restores stability and improves reconstruction fidelity.
The discrepancy becomes most dramatic for chirp-pulsed excitation (Fig.~\ref{fig5}(c)). While the subspace-sampled approach continues to yield consistent virtual-wave reconstruction and well-localized defect signatures, the original inversion fails catastrophically: the reconstructed signals exhibit strong distortion, poor agreement with the forward model, and highly degraded or spurious imaging results. This breakdown reflects the intrinsically high-dimensional and broadband spectral sampling of chirped excitation. In the absence of subspace constraints, the inversion amplifies the ill-posedness associated with diffusion-induced spectral attenuation, leading to severe instability and noise amplification. By contrast, the subspace framework effectively regularizes the problem by confining the solution to an excitation-induced low-dimensional manifold, thereby maintaining reconstruction consistency even under broadband excitation.

\section{Conclusion}
We have developed a unified theoretical and computational framework for virtual-wave reconstruction based on a generalized operator formulation of diffusion–wave mapping. By casting the transformation between thermal diffusion and virtual wave fields as a causal Fredholm operator with analytic continuation in the complex frequency domain, we provide a rigorous foundation that bridges previously disparate formulations and excitation-specific approaches.
A central result of this work is the identification of excitation-induced subspace sampling as the key mechanism governing reconstruction behavior. Within this perspective, commonly used schemes—pulse, lock-in, chirped, and coded excitations—are not fundamentally different transforms, but rather distinct projections or sampling strategies of a single generalized operator. This interpretation establishes a direct link between excitation design, spectral observability, and the conditioning of the inverse problem.
Through both simulations and experiments, we demonstrate that subspace-constrained inversion acts as an intrinsic regularization mechanism. While pulsed excitation already approximates full-space sampling and yields comparable results with or without subspace constraints, structured excitations such as lock-in and chirped signals reveal a clear breakdown of the unconstrained formulation. In these cases, the original inversion becomes increasingly ill-posed, exhibiting amplitude distortion, loss of coherence, and ultimately catastrophic failure for broadband excitation. In contrast, subspace sampling systematically restores stability by confining the solution to a physically admissible low-dimensional manifold, ensuring robust reconstruction and consistent defect imaging across excitation regimes.
These findings establish a unified spectral–geometric interpretation of virtual-wave imaging, in which reconstruction quality is governed by the interplay between subspace dimensionality and diffusion-induced information loss. More broadly, the framework provides a principled route for excitation optimization and inverse-problem design, enabling controlled trade-offs between stability and information capacity. This opens new perspectives for high-resolution thermographic imaging, photothermal tomography, and, more generally, the analysis of irreversible transport processes through wave-based representations.

\begin{acknowledgments}
	This work was supported by the Adolf Martens Fellowship (Grant n. BAM-AMF-2025-1).
\end{acknowledgments}

\appendix
\section{A unified generalized virtual wave operator}
Building upon the kernel representation in Eq.~\eqref{eq11}, the mapping between the virtual wave field and the diffusion field can be recast in an operator form. Specifically, we define a linear integral operator $\mathcal{V}$, referred to as the generalized virtual wave transform (GVWT), as
\begin{equation}
	T(\mathbf{r}, t) = \mathcal{V}[T_{\mathrm{virt}}](\mathbf{r}, t)
	= \int_{0}^{\infty}
	 K(t, t') \, T_{\mathrm{virt}}(\mathbf{r}, t')\, dt',
	\label{a1}
	\end{equation}
where the kernel $K(t, t')$ is given by Eq.~\eqref{eq12}. The operator $\mathcal{V}$ maps a propagating virtual wave field into a diffusive temperature field, thereby establishing a unified transformation framework between wave-like and diffusion-like dynamics.

The operator $\mathcal{V}$ possesses several important structural properties: (i) Linearity. The mapping is linear with respect to the virtual wave field,
\begin{equation}
	\mathcal{V}[aT_{1}+bT_{2}]=a\mathcal{V}[T_{1}]+b\mathcal{V}[T_{2}],
	\label{a2}
\end{equation}
which directly follows from the linearity of the governing equations. (ii) Causality. The kernel satisfies $K(t, t')=0$ for $t<0$ or $t'<0$, ensuring that $T(\mathbf{r}, t)$ depends only on past contributions of the virtual wave field. This reflects the irreversible nature of thermal diffusion. (iii) Smoothing (diffusive filtering). The exponential factor in Eq.~\eqref{eq12}, $\exp\left(-{c^2 t'^2}/{4\alpha t}\right)$, acts as a diffusion-induced low-pass filter, suppressing high-frequency components of the virtual wave field. As a result, $\mathcal{V}$ is a compact operator that inherently leads to information loss, which is consistent with the ill-posed nature of inverse heat conduction problems. (iv) Scale coupling between $t$ and $t'$. Unlike standard convolution operators, the kernel $K(t, t')$ is not shift-invariant, it couples $t$ and $t'$ through the ratio $t'/\sqrt{t}$, reflecting the non-stationary nature of diffusion-wave propagation.

The operator $\mathcal{V}$ admits a compact representation in the frequency domain. From Eq.~\eqref{eq8}, the transformation can be expressed as
\begin{equation}
	\tilde{T}(\mathbf{r}, \omega)
	= \frac{c^2}{\kappa}\,
	\tilde{T}_{\mathrm{virt}}(\mathbf{r}, -ic\sigma(\omega)),
	\label{a3}
\end{equation}
which can be interpreted as an analytic continuation in the complex frequency plane. In this sense, the GVWT can be interpreted as a spectral mapping that converts oscillatory wave solutions into exponentially decaying diffusion modes. In practice, the virtual wave field is not arbitrary but is generated by a prescribed excitation waveform. Let the source term be written as
\begin{equation}
	Q(\mathbf{r},t)=q(\mathbf{r})s(t),
	\label{a4}
\end{equation}
where $s(t)$ denotes the temporal excitation profile. Due to linearity, the virtual wave field can be expressed as
\begin{equation}
	T_{\mathrm{virt}}(\mathbf{r}, t)
	=
	\int_{0}^{\infty}
	G_{\mathrm{wave}}(\mathbf{r}, t-\tau)\, s(\tau)\, d\tau,
	\label{a5}
\end{equation}
where $G_\mathrm{wave}$ is the Green's function of the wave equation. Substituting Eq.~\eqref{eq17} into Eq.~\eqref{eq13}, the temperature field becomes
\begin{equation}
	T(\mathbf{r}, t)=
	\int_{0}^{\infty} K(t,t')
	\left[
	\int_{0}^{\infty} G_{\mathrm{wave}}(t'-\tau)\, s(\tau)\, d\tau
	\right]
	dt'
	\label{a6}
\end{equation}

Assuming absolute integrability, the order of integration can be exchanged, yielding
\begin{equation}
	T(\mathbf{r}, t)=
	\int_{0}^{\infty} \mathcal{K}_{s}\left(t,\tau\right)s(\tau)\, d\tau,
	\label{a7}
\end{equation}
which defines an excitation-dependent effective operator
\begin{equation}
	\mathcal{K}_{s}(t,\tau)=
	\int_{0}^{\infty} K\left(t,t'\right)G_{\mathrm{wave}}(\mathbf{r}, t'-\tau)dt',
	\label{a8}
\end{equation}

This formulation reveals that the overall mapping from the excitation $s(t)$ to the temperature field $T(\mathbf{r},t)$ can be decomposed into a composition of two linear operators:
\begin{equation}
	s(t)
	\;\xrightarrow{\;\mathcal{W}\;}\;
	T_{\mathrm{virt}}(\mathbf{r},t)
	\;\xrightarrow{\;\mathcal{V}\;}\;
	T(\mathbf{r},t),
	\label{a9}
\end{equation}
where $\mathcal{W}$ denotes the wave propagation operator determined by the wave Green's function,
\begin{equation}
	T_{\mathrm{virt}}(\mathbf{r},t)
	=
	\mathcal{W}[s]
	=
	\int_{0}^{\infty}
	G_{\mathrm{wave}}(\mathbf{r},t-\tau)\,
	s(\tau)\,d\tau,
	\label{a10}
\end{equation}
$\mathcal{V}$ denotes the generalized virtual wave transform (GVWT),
\begin{equation}
	T(\mathbf{r},t)
	=
	\mathcal{V}[T_{\mathrm{virt}}]
	=
	\int_{0}^{\infty}
	K(t,t')\,
	T_{\mathrm{virt}}(\mathbf{r},t')\,dt'.
	\label{a11}
\end{equation}

Therefore, excitation-dependent measurements are governed by the composite operator
\begin{equation}
	\mathcal{A}
	=
	\mathcal{V}
	\circ
	\mathcal{W},
	\label{a12}
\end{equation}
rather than by $\mathcal{V}$ alone.

\section{Projection of the Generalized Virtual Wave Transform under Modulated Excitations}
With the generalized framework established in the previous section, different virtual wave transform schemes arise naturally from the choice of excitation waveform. In this section, we demonstrate that commonly used excitation strategies, including Dirac-pulsed, lock-in, chirped, and coded excitations, can be interpreted as distinct degeneracy limits of the same operator $\mathcal{V}$. Rather than representing fundamentally different transforms, these methods correspond to different projections of the virtual wave field onto excitation-dependent subspaces.

The Dirac pulse excitation represents a fundamental limiting case of the generalized virtual wave transform (GVWT), providing direct access to the underlying operator kernel without excitation-induced projection or compression. For an impulsive excitation
For an impulsive excitation 
\begin{equation}
 s(t) = \delta(t),
 	\label{b1}
\end{equation}
with
\begin{equation}
Q(\mathbf{r},t) = q(\mathbf{r})\delta(t),
 	\label{b2}
\end{equation}

The virtual wave field reduces to the Green’s function of the wave equation,
\begin{equation}
	T_{\mathrm{virt}}(\mathbf{r},t) = G_{\mathrm{wave}}(\mathbf{r},t).
	 	\label{b3}
\end{equation}

The resulting temperature field is then given by
\begin{equation}
	T(\mathbf{r},t) = \int_{0}^{\infty} K(t,t')\, G_{\mathrm{wave}}(\mathbf{r},t') \, dt'.
	 	\label{b4}
\end{equation}

Equivalently, within the excitation–operator representation,
\begin{equation}
	T(\mathbf{r},t) = \int_{0}^{\infty} K_s(t,\tau)\, s(\tau)\, d\tau,
	 	\label{b5}
\end{equation}
one obtains the simplified form
\begin{equation}
	T(\mathbf{r},t) = K_s(t,0).
	 	\label{b6}
\end{equation}

For monochromatic excitation,
\begin{equation}
	s(t) = e^{i\omega_{0} t},
		\label{b7}
\end{equation}
the virtual wave field is generated through the wave propagation operator,
\begin{equation}
	T_{\mathrm{virt}} = \mathcal{W}\!\left[e^{i\omega_{0} t}\right],
		\label{b8}
\end{equation}
and the measured diffusion field becomes
\begin{equation}
	T(\mathbf{r},t)
	=
	(\mathcal{V} \circ \mathcal{W})
	\!\left[
	e^{i\omega_{0} t}
	\right].
		\label{b9}
\end{equation}

Equivalently,
\begin{equation}
	T(\mathbf{r},t)
	=
	\int_{0}^{\infty}
	K_{s}(t,\tau)\,
	e^{i\omega_{0} \tau}\,
	d\tau.
		\label{b10}
\end{equation}
This shows that lock-in detection corresponds to a rank-1 projection of the excitation space, where only a single Fourier mode of the virtual wave field contributes to the reconstructed diffusion field. Physically, this leads to strong phase selectivity but limited temporal resolution, as all depth information is encoded into a single harmonic response.

For chirped excitation,
\begin{equation}
	s(t) = e^{i(\omega_{0} t + \beta t^{2})},
		\label{b11}
\end{equation}
the excitation first generates a virtual wave field through the wave propagation operator,
\begin{equation}
	T_{\mathrm{virt}} = \mathcal{W}[s],
		\label{b12}
\end{equation}
and the measured temperature field is given by the composite operator
\begin{equation}
	T(\mathbf{r},t)
	=
	(\mathcal{V} \circ \mathcal{W})[s].
		\label{b13}
\end{equation}

Equivalently, the response can be expressed in excitation-space representation as
\begin{equation}
	T(\mathbf{r},t)
	=
	\int_{0}^{\infty}
	\mathcal{K}(t,\tau)\,
	e^{i(\omega_{0}\tau + \beta \tau^{2})}
	d\tau,
		\label{b14}
\end{equation}
where $\mathcal{K}(t,\tau)$ is the excitation-independent effective kernel defined by the composition of $\mathcal{V}$ and $\mathcal{W}$.

From a time–frequency perspective, the chirped excitation introduces a continuously varying instantaneous frequency
\begin{equation}
	\omega_{\mathrm{inst}}(\tau)
	=
	\omega_{0} + 2\beta \tau,
		\label{b15}
\end{equation}
which implies that the excitation does not probe a single spectral mode, but rather traces a continuous trajectory in frequency space.

In this sense, the integral can be interpreted as a continuous superposition of kernel responses weighted along a time–frequency path. This provides an intuitive interpretation of chirped excitation as a structured sampling strategy of the composite operator $(\mathcal{V}\circ\mathcal{W})$, rather than a single-frequency projection as in lock-in excitation. Importantly, this interpretation is geometric rather than asymptotic: it does not rely on stationary phase approximation, but instead follows directly from the intrinsic time–frequency structure of the excitation waveform.

For coded excitation, the source is constructed as a superposition of discrete symbols,
\begin{equation}
	s(t) = \sum_{n}^{} a_{n} \phi_{n}(t),
	\label{b16}
\end{equation}
where ${\phi_{n}(t)}$\(x\) represents an orthogonal coding basis (e.g., M-sequence, Hadamard, or pseudo-random binary sequences). Substituting Eq.~\eqref{b16} into Eq.~\eqref{a7}, we obtain
\begin{equation}
	T_{\mathrm{coded}}(\mathbf{r},t) = 
	\sum_{n}^{} a_{n} \int_{0}^{\infty} \mathcal{K}(t,t')\phi_{n}(t')dt'.
	\label{b17}
\end{equation}

This formulation reveals that coded excitation corresponds to a finite-dimensional projection of the operator onto an orthogonal temporal basis. Unlike lock-in and chirped schemes, coded excitation enables parallel multiplexing of temporal modes, effectively increasing the information capacity of the measurement system. In the Appendix~\ref{app:subsecA2} and Appendix~\ref{app:subsecA3}, we show that several commonly used schemes, including chirped pulse excitation and pseudo-noise (PN) sequences, can be interpreted as special cases of this coded excitation model under specific choices of the basis functions $\{\phi_n(t)\}$.

Eqs.~\eqref{b2},~\eqref{b7}, together with the chirped pulse (Eq.~\eqref{b11}) and PN-sequence formulations (Eq.~\eqref{b16}), demonstrate that all excitation schemes can be interpreted as different degeneracy limits or sampling strategies of the same generalized linear operator $\mathcal{V}$.

In the most general form, the excitation-to-temperature mapping is governed by the composite operator
\begin{equation}
	\mathcal{A} = \mathcal{V} \circ \mathcal{W},
	\label{b18}
\end{equation}
such that
\begin{equation}
	T(\mathbf{r},t)
	=
	\mathcal{A}[s](\mathbf{r},t)
	=
	\int_{0}^{\infty}
	K_{s}(t,\tau)\,s(\tau)\,d\tau.
	\label{b19}
\end{equation}

Here, $K_{s}(t,\tau)$ denotes the effective excitation-dependent kernel obtained from the composition of wave propagation and diffusion transformation. (i) Lock-in excitation (rank-1 projection). For monochromatic excitation,
\begin{equation}
	s(t') = e^{i\omega t'},
	\label{b20}
\end{equation}
the operator reduces to a single-mode projection,
\begin{equation}
	T(t) = \langle K(t,\cdot), e^{i\omega t'} \rangle,
	\label{b21}
\end{equation}
corresponding to a rank-1 sampling of the operator spectrum. (ii) Continuous chirped excitation (spectral trajectory). For ideal chirped excitation,
\begin{equation}
	s(t') = e^{i(\omega_0 t' + \beta t'^2)},
	\label{b22}
\end{equation}
the instantaneous frequency
\begin{equation}
	\omega_{\mathrm{inst}}(t') = \omega_0 + 2\beta t'
	\label{b23}
\end{equation}
defines a continuous spectral trajectory, probing the operator along a one-dimensional manifold in frequency space. (iii) Coded excitation (finite-dimensional subspace).
For general coded excitation,
\begin{equation}
	s(t') = \sum_{m} a_m \phi_m(t'),
	\label{b24}
\end{equation}
the response becomes
\begin{equation}
	T(t) = \sum_{m} a_m \langle K(t,\cdot), \phi_m(t') \rangle,
	\label{b25}
\end{equation}
which corresponds to a projection onto a finite-dimensional subspace. (iv) Linear frequency-modulated (LFM) pulse excitation (time-localized trajectory). For LFM Gaussian pulse trains,
\begin{equation}
	s(t')
	=
	\sum_{n}
	I_0 \exp\!\left(-\frac{(t'-t_n)^2}{\xi}\right),
	\label{b26}
\end{equation}
the operator response becomes
\begin{equation}
	T(t)
	=
	\sum_{n}
	\int_{0}^{\infty}
	K(t,t')\,
	I_0 \exp\!\left(-\frac{(t'-t_n)^2}{\xi}\right)
	dt'.
	\label{b27}
\end{equation}

This corresponds to a localized sampling of the excitation function, where each pulse probes a neighborhood of $t'=t_n$. Compared with continuous chirp, LFM excitation introduces a finite temporal bandwidth, resulting in a Gaussian-smoothed sampling of the operator. (v) PN-sequence excitation (discrete sampling of kernel shifts). For pseudo-noise (PN) excitation,
\begin{equation}
	s(t') = \sum_{n} s[n]\,\delta(t'-n\Delta t),
	\label{b28}
\end{equation}
the operator reduces to
\begin{equation}
	T(t)
	=
	\sum_{n}
	s[n]\,
	K(t, n\Delta t),
	\label{b29}
\end{equation}
which corresponds to direct discrete sampling of the kernel along the $t'$ dimension. Thus, PN excitation performs a coded sampling of the operator by selecting shifted kernel components.

All excitation schemes can be interpreted as selecting different subspaces or sampling strategies in the excitation function space:
\begin{equation}
	s(t') \in \mathcal{S} \subset L^2(\mathbb{R}^+),
	\label{b30}
\end{equation}
with the following hierarchy: (i) Lock-in: single-point sampling (rank-1); (ii) Continuous chirp: continuous spectral trajectory; (iii) Coded excitation: finite-dimensional subspace projection; (iv) LFM pulses: localized (windowed) trajectory sampling;  (v) PN sequences: discrete sampling of kernel shifts.

Therefore, excitation design can be interpreted as selecting different sampling strategies of the operator $\mathcal{V}$, providing a unified framework for analyzing resolution, bandwidth, and information capacity in diffusion–wave mapping systems.

\section{Excitation-Induced Subspace Reduction and Dual-Projection Structure of GVWT}
The ill-posedness of the generalized virtual wave transform (GVWT) is intrinsically related to the dimensionality of the admissible virtual-wave solution space. In the unrestricted case, the virtual wave field
\begin{equation}
	T_{\mathrm{virt}}(\mathbf r,t)
	\in
	L^2(\mathbb R^+)
\end{equation}
contains infinitely many temporal degrees of freedom, and the inversion of the diffusion operator
\begin{equation}
	T(\mathbf r,t)
	=
	\int_0^\infty
	K(t,t')T_{\mathrm{virt}}(\mathbf r,t')dt'
\end{equation}
constitutes a severely ill-posed infinite-dimensional inverse problem.

However, under prescribed excitation, the virtual wave field is no longer arbitrary. Since
\begin{equation}
	T_{\mathrm{virt}}(\mathbf r,t)
	=
	\int_0^\infty
	G_{\mathrm{wave}}(\mathbf r,t-\tau)
	s(\tau)d\tau,
\end{equation}
the excitation waveform $s(t)$ constrains the admissible temporal modes of the virtual wave field. Consequently, the inversion is effectively restricted to an excitation-induced subspace
\begin{equation}
	\mathcal H_s
	\subset
	L^2(\mathbb R^+),
\end{equation}
whose dimensionality is determined by the structure of the excitation.

Therefore, the practical difficulty of the inverse problem is no longer governed solely by the compactness of the diffusion operator, but also by the dimension of the excitation-induced subspace. In this sense, excitation design acts as a dimensionality-control mechanism for the inverse problem.

For monochromatic lock-in excitation,
\begin{equation}
	s(t)=e^{i\omega_0 t},
\end{equation}
the admissible virtual-wave space collapses into the one-dimensional harmonic subspace
\begin{equation}
	\mathcal H_{\omega_0}
	=
	\mathrm{span}{e^{i\omega_0 t}}.
\end{equation}
The inversion therefore reduces to estimating a single complex modal coefficient, leading to strong stability but limited information capacity.

For chirped excitation,
\begin{equation}
	s(t)=e^{i(\omega_0 t+\beta t^2)},
\end{equation}
the admissible solutions are constrained to a continuous frequency-manifold subspace parameterized by the instantaneous frequency trajectory,
\begin{equation}
	\omega_{\mathrm{inst}}(t)=\omega_0+2\beta t.
\end{equation}
Compared with lock-in excitation, the effective subspace dimension increases, enabling improved spectral coverage at the expense of reduced conditioning.

For coded excitation,
\begin{equation}
	s(t)=\sum_m a_m\phi_m(t),
\end{equation}
the admissible virtual-wave field belongs to the finite-dimensional span
\begin{equation}
	\mathcal H_{\mathrm{coded}}
	=
	\mathrm{span}{\phi_m(t)}_{m=1}^{M},
\end{equation}
whose dimension is determined by the number of coding modes $M$. Increasing the number of basis functions improves information capacity while simultaneously increasing inversion sensitivity.

For PN-sequence excitation,
\begin{equation}
	s(t)=\sum_n s[n]\delta(t-n\Delta t),
\end{equation}
the virtual wave field becomes a discrete superposition of shifted Green's functions,
\begin{equation}
	T_{\mathrm{virt}}(t)
	=
	\sum_n s[n]
	G_{\mathrm{wave}}(t-n\Delta t),
\end{equation}
corresponding to a high-dimensional discrete sampling subspace. Such excitation provides broad spectral coverage and high information capacity, but also leads to stronger sensitivity to noise and model uncertainty.

These observations reveal a general trade-off between inversion stability and information capacity:
\begin{equation}
C_{\mathrm{inv}} \sim f\!\left(\dim(\mathcal{H}_{s})\right)
\end{equation}
where $\mathcal{C}_{\mathrm{inv}}$ denotes the effective complexity (or conditioning difficulty) of the inverse problem, and $\dim\!\left(\mathcal{H}_{s}\right)$ denotes the dimensionality of the excitation-induced subspace. Low-dimensional excitation subspaces provide strong regularization but limited recoverable information, whereas high-dimensional subspaces increase reconstruction capability at the cost of stronger ill-posedness.

Therefore, excitation design can be interpreted as a mechanism for controlling the effective dimensionality, conditioning, and information content of the generalized virtual wave inverse problem.

In practical implementations, the reconstruction of the virtual wave field from temperature measurements is formulated as a regularized inverse problem. Starting from the composite operator relation
\begin{equation}
	T(\mathbf{r}, t)
	=
	\mathcal{V}\circ\mathcal{W}[s]
	=
	\int_{0}^{\infty}
	K_s(t,t')\, s(t')\, dt',
\end{equation}
we assume that the virtual wave field is constrained within the excitation-induced subspace $\mathcal{H}_s$.

To make this constraint explicit, we introduce a regularized inversion formulation:
\begin{equation}
	\min_{T_{\mathrm{virt}} \in \mathcal{H}_s}
	\left\|
	T - \mathcal{V}[T_{\mathrm{virt}}]
	\right\|_{L^2}^2
	+
	\lambda \, \mathcal{R}[T_{\mathrm{virt}}],
	\label{eqA200}
\end{equation}
where $\mathcal{R}[\cdot]$ is a stabilizing regularization functional and $\lambda > 0$ controls the trade-off between data fidelity and stability.

Using the excitation-induced basis expansion
\begin{equation}
	T_{\mathrm{virt}}(\mathbf{r},t)
	=
	\sum_{k=1}^{M} a_k(\mathbf{r}) \, \psi_k^{(s)}(t),
\end{equation}
the variational problem reduces to a finite-dimensional optimization:
\begin{equation}
	\min_{\mathbf{a}(\mathbf{r})}
	\left\|
	T(\mathbf{r},t)
	-
	\sum_{k=1}^{M} a_k(\mathbf{r}) \,
	\mathcal{V}[\psi_k^{(s)}](t)
	\right\|_{L^2}^2
	+
	\lambda \|\mathbf{a}(\mathbf{r})\|_2^2.
	\label{eqA201}
\end{equation}

Taking the first-order optimality condition yields the normal equation:
\begin{equation}
	\left(\mathbf{A}^{(s)\top}\mathbf{A}^{(s)} + \lambda \mathbf{I}\right)\mathbf{a}
	=
	\mathbf{A}^{(s)\top}\mathbf{T},
	\label{eqA202}
\end{equation}
where $\mathbf{A}^{(s)}$ is the excitation-dependent system matrix defined by
\begin{equation}
	A^{(s)}_{ik}
	=
	\left[\mathcal{V}[\psi_k^{(s)}]\right](t_i).
\end{equation}

Finally, the reconstructed virtual wave field is obtained as
\begin{equation}
	T_{\mathrm{virt}}(\mathbf{r},t)
	=
	\sum_{k=1}^{M}
	a_k(\mathbf{r}) \, \psi_k^{(s)}(t),
\end{equation}
which is guaranteed to lie within the excitation-induced subspace $\mathcal{H}_s$ by construction.

This formulation shows that the inverse GVWT problem is not solved in an unconstrained function space, but rather within a regularized, excitation-dependent low-dimensional manifold, where the ill-posedness is controlled by both the regularization parameter $\lambda$ and the effective subspace dimension $M$.

\section{\label{app:subsecA1}Rigorous justification of the analytic continuation}
To justify the mapping
\begin{equation}
	\tilde{T}(\mathbf{r}, \omega)
	= \frac{c^2}{\kappa}\,
	\tilde{T}_{\mathrm{virt}}(\mathbf{r}, -ic\sigma(\omega)),
	\label{eqA1}
\end{equation}
We interpret it as an analytic continuation of the temporal Fourier transform of the virtual wave field into the complex frequency plane. Let the virtual wave field satisfy 
\begin{equation}
T_{\mathrm{virt}}(\mathbf{r},t)\in L^{1}(\mathbb{R}_+)
\label{eqA2}
\end{equation}
with at most exponential growth, i.e.
\begin{equation}
\left|T_{\mathrm{virt}}(\mathbf{r},t)\right|\le C e^{a t}, \quad a>0.
\label{eqA3}
\end{equation}

Its Laplace–Fourier transform is defined as
\begin{equation}
\tilde{T}_{\mathrm{virt}}(\mathbf{r},z)
=
\int_{0}^{\infty} T_{\mathrm{virt}}(\mathbf{r},t)\,e^{-izt}\,dt,
\label{eqA4}
\end{equation}
which defines a holomorphic function in the half-plane
\begin{equation}
\Im z < a.
\label{eqA5}
\end{equation}

The diffusion wavenumber is defined as
\begin{equation}
\sigma(\omega)=\sqrt{\frac{i\omega}{\alpha}},
\label{eqA6}
\end{equation}
where the principal branch of the square root is chosen such that
\begin{equation}
\arg(i\omega)\in(-\pi,\pi), \qquad \Re\,\sigma(\omega)>0,
\label{eqA7}
\end{equation}
ensuring spatial decay of diffusion modes.

Under this choice, the complex map
\begin{equation}
z(\omega)=-ic\,\sigma(\omega)
\label{eqA8}
\end{equation}
satisfies
\begin{equation}
\Im z(\omega)=-c\,\Re \sigma(\omega)<0
\quad \text{for all real } \omega,
\label{eqA9}
\end{equation}
implying that the image of the real frequency axis lies entirely within the domain of analyticity of \(\tilde{T}_{\mathrm{virt}}(\mathbf{r},z)\). Therefore, the evaluation
\begin{equation}
\tilde{T}_{\mathrm{virt}}(\mathbf{r},-ic\sigma(\omega))
\label{eqA10}
\end{equation}
is well defined as an analytic continuation. Moreover, by the uniqueness of holomorphic extensions, this continuation is uniquely determined by the values of \(\tilde{T}_{\mathrm{virt}}\) on the real axis. The transformation thus defines a composition of a complex spectral reparameterization and a Laplace-analytic extension, rather than an ad hoc substitution.

\section{\label{app:subsecA2}Linear Frequency-Modulated Chirped Pulse Excitation}
From the perspective of Eq.~\eqref{b16}, the linear frequency-modulated (LFM) chirped pulse excitation can be interpreted as a coded representation of a chirped waveform, expressed as a sequence of temporally shifted Gaussian pulses. The excitation is given by
\begin{equation}
	I_{\mathrm{exc}}(t)
	=
	\sum_{n=0}^{p-1}
	I_0 \exp\!\left[-\frac{(t - t_n)^2}{\xi}\right],
\end{equation}
where the non-uniform sampling times $\{t_n\}$ are determined by the chirp phase law.

This representation can be equivalently written as a convolution between a Gaussian envelope and a discrete temporal sampling sequence,
\begin{equation}
	I_{\mathrm{exc}}(t)
	=
	I(t) \otimes
	\sum_{n=0}^{p-1}
	\delta(t - t_n),
\end{equation}
where
\begin{equation}
	I(t)=I_0 \exp\!\left(-\frac{t^2}{\xi}\right).
\end{equation}

In the frequency domain, the excitation separates into a spectral envelope and a discrete phase coding term,
\begin{equation}
	\tilde{I}_{\mathrm{exc}}(\omega)
	=
	\tilde{I}(\omega)
	\sum_{n=0}^{p-1}
	e^{-i\omega t_n},
\end{equation}
where
\begin{equation}
	\tilde{I}(\omega)
	=
	I_0\sqrt{\pi\xi}\exp\!\left(-\frac{\xi\omega^2}{4}\right).
\end{equation}

The diffusion field is obtained via the generalized virtual wave transform applied to the wave-propagated excitation, i.e.,
\begin{equation}
	T(\mathbf r,t)
	=
	(\mathcal V \circ \mathcal W)[s].
\end{equation}

Substituting the spectral representation of the excitation leads to a superposition form in which each chirp component contributes a time-shifted response of the composite operator. Transforming back to the time domain yields a kernel representation of the form
\begin{equation}
	T(\mathbf r,t)
	=
	\sum_n
	\int_0^\infty
	\mathcal{K}(t,\tau)\,
	\delta(\tau - t_n)
	d\tau,
\end{equation}
which reflects the discrete sampling nature of the chirped pulse train.

The corresponding effective kernel can be interpreted as a superposition of time-shifted diffusion responses weighted by a Gaussian smoothing envelope,
\begin{equation}
	K_{\mathrm{chirp}}(t,t')
	=
	\sum_{n}
	\int_{t_n}^{t}
	K(\tau,t')\,
	G_{\xi}(t-\tau)
	d\tau,
\end{equation}
where
\begin{equation}
	G_{\xi}(t)
	=
	\frac{1}{\sqrt{\pi\xi}}
	\exp\!\left(-\frac{t^2}{\xi}\right).
\end{equation}

This representation shows that the LFM chirped pulse excitation does not modify the generalized virtual wave transform kernel itself, but instead induces a structured weighting of the excitation-dependent projection of the composite operator $(\mathcal{V}\circ\mathcal{W})$. In this sense, the chirped pulse train corresponds to a non-uniform sampling of the excitation subspace $\mathcal{S}$, where the sampling locations $\{t_n\}$ are determined by the instantaneous frequency law of the chirp.

\section{\label{app:subsecA3}Pseudo-Noise Sequence Excitation}
Rearranging Eq.~\eqref{b17}, the PN-induced response can be written in a compact operator form:
\begin{equation}
	T_{\mathrm{PN}}(\mathbf{r}, t)
	=
	\sum_{n=0}^{N-1}
	s[n]
	\int_{0}^{\infty}
	K(t, t')\,
	G_{\mathrm{wave}}(\mathbf{r}, t' - n\Delta t)\, dt'.
	\label{eq53_new}
\end{equation}

This expression shows that pseudo-noise excitation does not modify the GVWT kernel itself, but instead induces a discrete superposition of time-shifted Green’s functions in the excitation space.

Under the time-invariance of the wave Green’s function, i.e.,
$G_{\mathrm{wave}}(t,t') = G_{\mathrm{wave}}(t - t')$, the temporal shift induced by the PN sequence can be equivalently transferred to the kernel representation.
\begin{equation}
	T_{\mathrm{PN}}(\mathbf{r}, t)
	=
	\int_{0}^{\infty}
	T_{\mathrm{virt}}(\mathbf{r}, t')
	K_{\mathrm{PN}}(t, t')\, dt',
	\label{eq54_new}
\end{equation}

where the PN-modulated kernel is defined as
\begin{equation}
	K_{\mathrm{PN}}(t, t')
	=
	\sum_{n=0}^{N-1}
	s[n]\,
	K(t, t' - n\Delta t).
	\label{eq55_new}
\end{equation}
This equivalence follows directly from the convolution structure of the wave propagation operator. For Maximum Length Sequences,
\begin{equation}
	s_{\mathrm{MLS}}[n] \in \{ \pm 1 \}, \quad N = 2^{M} - 1,
	\label{eq56}
\end{equation}
the kernel becomes
\begin{equation}
	K_{\mathrm{MLS}}(t, t')
	=
	\sum_{n=0}^{2^{M}-2}
	s_{\mathrm{MLS}}[n]\,
	K(t, t' - n\Delta t).
	\label{eq57}
\end{equation}

MLS approximates a flat-spectrum excitation, enabling near-uniform sampling of the operator spectrum.

For Legendre sequences,
\begin{equation}
	s_{\mathrm{LS}}[n] = \left( \frac{n}{p} \right), \quad p \ \text{prime},
	\label{eq58}
\end{equation}
the kernel becomes
\begin{equation}
	K_{\mathrm{LS}}(t, t')
	=
	\sum_{n=0}^{p-1}
	\left( \frac{n}{p} \right)
	K(t, t' - n\Delta t).
	\label{eq59}
\end{equation}

Legendre sequences introduce deterministic phase-coded modulation with structured spectral correlations.

Taking the Fourier transform with respect to $t'$, we obtain
\begin{equation}
	\tilde{K}_{\mathrm{PN}}(t,\omega)
	=
	\tilde{K}(t,\omega)\, S(\omega),
	\label{eq60}
\end{equation}
where
\begin{equation}
	S(\omega)
	=
	\sum_{n}
	s[n]\,
	e^{-i\omega n \Delta t}.
	\label{eq61}
\end{equation}

Thus, PN excitation acts as a spectral shaping operator on the virtual wave kernel. Combining results, the PN-based virtual wave transform can be written compactly as
\begin{equation}
	T(\mathbf{r}, t)
	=
	\int_{0}^{\infty}
	T_{\mathrm{virt}}(\mathbf{r}, t')\,
	K_{\mathrm{PN}}(t, t')\, dt',
	\label{eq62}
\end{equation}
with
\begin{equation}
	K_{\mathrm{PN}}(t, t')
	=
	\sum_{n=0}^{N-1}
	s[n]\,
	K(t, t' - n\Delta t).
	\label{eq63}
\end{equation}

Therefore, PN excitation corresponds to a discrete spectral weighting of the GVWT kernel, where the sequence $s[n]$ defines a deterministic sampling measure in the excitation subspace. Unlike continuous chirped excitation, PN sequences provide non-parametric and fully discrete access to the operator spectrum.

\end{document}